\documentclass[12pt]{article}
\oddsidemargin = \evensidemargin
\newfont{\sase}{cmss12 scaled 1300}
\newfont{\sasi}{cmss12 scaled 1500}
\usepackage{colordvi}
\usepackage{amssymb}
\usepackage[american]{babel}

\newcommand{\sgn}{\mathrm{sgn}\,}
\newcommand{\enquote}[1]{``#1''}

\begin{document}

\title{\bf Contextual Emergence of Mental States from Neurodynamics}

\author{Harald Atmanspacher \\
Institute for Frontier Areas of Psychology and Mental Health, \\
Freiburg, Germany \\
and \\
Parmenides Foundation, \\
Capoliveri, Italy
\and
Peter beim Graben \\
Institute of Linguistics and Institute of Physics, \\
University of Potsdam, Germany
}

\date{\today}
\maketitle

\bigskip

\begin{abstract}

\noindent The emergence of mental states from neural states by partitioning the neural phase space is
analyzed in terms of symbolic dynamics. Well-defined mental states provide contexts inducing a criterion of
structural stability for the neurodynamics that can be implemented by particular partitions. This leads to
distinguished subshifts of finite type that are either cyclic or irreducible. Cyclic shifts correspond to
asymptotically stable fixed points or limit tori whereas irreducible shifts are obtained from generating
partitions of mixing hyperbolic systems. These stability criteria are applied to the discussion of neural
correlates of consiousness, to the definition of macroscopic neural states, and to aspects of the symbol
grounding problem. In particular, it is shown that compatible mental descriptions, topologically equivalent
to the neurodynamical description, emerge if the partition of the neural phase space is generating. If this
is not the case, mental descriptions are incompatible or complementary. Consequences of this result for an
integration or unification of cognitive science or psychology, respectively, will be indicated.

\end{abstract}

\section{Interlevel Relations}

Knowledge of well-defined relations among different levels of descriptions of physical and other systems is
inevitable if one wants to understand how (elements of) different descriptions depend on each other, give
rise to each other, or even imply each other. The most ambitious program in this regard is physical reduction
in the sense that higher-level descriptions of features of a system are determined by the description of
features at the most fundamental level of physical theory, no matter how remote the higher level is from that
most fundamental level. This program assumes that the description of all features which are not included at
the fundamental level can be constructed or derived from this level without additional input.

However, already physical examples pose serious difficulties for this program. It has recently been proposed
that the concept of {\it contextual emergence} (Atmanspacher and Bishop, this issue; Bishop and Atmanspacher,
preprint) addresses such situations more properly. Contextual emergence is characterized by the fact that
lower-level descriptions provide necessary, but not sufficient conditions for higher-level descriptions.
(Note that such a relation between descriptive levels does not necessarily entail the same relation between
ontological levels.) The presence of necessary conditions indicates that lower-level descriptions provide a
basis for higher-level descriptions, while the absence of sufficient conditions means that higher-level
features are neither logical consequences of lower-level descriptions nor can they be rigorously derived from
them alone. Hence, a full-blown reductive program is inapplicable in these cases. Sufficient conditions for a
rigorous derivation of higher-level features can be introduced through specifying contexts reflecting the
particular kinds of contingency in a given situation.

A key ingredient of this procedure is the definition of some type of {\it stability condition} (e.g., the KMS
condition, due to Kubo, Martin, and Schwinger) based on considerations required to establish the framework of
a higher-level description (e.g., thermal equilibrium). This condition is often implemented as a reference
state with respect to which an asymptotic expansion is singular in the lower-level state space. Its
regularization defines a novel, contextual topology in which novel, emergent features can be rigorously
introduced. There is, thus, a mathematically well-defined procedure for deriving higher-level features given
the lower level description plus the contingent contextual conditions.

Contextual emergence and the associated identification of appropriate stability conditions may have
applications in other domains such as biology and psychology, and, ultimately, in the relationship between
the physical and the mental. In this contribution we will address a situation which is particularly difficult
because it exceeds the domain of material systems: relations between brain and consciousness. We will discuss
the contextual emergence of mental states and related features (psychology, cognitive science)  from brain
states and related features (neuroscience). Using Harnad's (1990) terms, this refers to the question of how
mental symbols and cognitive computation can be grounded in neurodynamics.

More specifically, mental representations will be considered as novel features at the (higher) level of
cognition, which have necessary but not sufficient conditions at the (lower) level of neuronal assemblies.
In order to identify contexts providing such sufficient conditions, those among the many possible cognitive
features that might be relevant or interesting as emergent features must first be identified. Assuming that
stability criteria play a role analogous to physical examples, techniques of modeling assemblies in terms of
generalized potentials with particular stability properties and corresponding relaxation times or escape
times suggest themselves. This can be implemented for powerful modeling tools such as neural networks
(Anderson and Rosenfeld 1989) or coupled map lattices (Kaneko and Tsuda 2000).

Other interlevel relations in addition to contextual emergence are strong reduction, radical emergence, and
supervenience (cf.~Atmanspacher and Bishop, this issue). While we do not think that strong reduction or
radical emergence provide clarifying insight for the relation between brain and consciousness, some comments
about supervenience are appropriate here. The notion of supervenience characterizes situations in which
lower-level descriptions contain sufficient but not necessary conditions for higher-level descriptions. This
scenario has been employed for brain-consciousness relations in the sense that a conscious state with a
particular phenomenal content can be multiply realized at the neural level (Kim 1992, 1993). For instance,
Chalmers (2000) defines neural correlates of consciousness (NCCs) as neural systems that are correlated with
conscious mental states and are minimally sufficient for the occurence of those states.

In this definition, the notion of sufficiency rather than necessity takes into account that different neural
states can be correlated with the same conscious state (multiple realization). Our notion of contextual
emergence addresses the different question of how neural states are related to conscious states in each
individual neural realization. Contextual emergence does not address the distinction between many-to-one and
one-to-one relations but tries to elucidate principles which allow us to understand the relationship between
mental and neural states itself, even in individual instantiations, in a more profound manner. In this way,
supervenience and contextual emergence complement rather than contradict each other. Applying both concepts
together may, thus, provide additional insight into the nature of mind-brain relations.

\section{Structural Stability in Symbolic Dynamics}

The issue of stability plays a prominent role in statistical mechanics. Haag {\it et al.} (1974) have shown
that Gibbs' thermal equilibrium states are uniquely characterized by three stability conditions upon state
functionals: (i) stationarity, i.e. expectation values of observables do not change in time; (ii) structural
stability, i.e., stability of the dynamics against perturbations; and (iii) ``asymptotic abelianness'', i.e.,
temporally distant observables become eventually compatible (see Bratteli and Robinson (1997)). From these
presuppositions, Haag {\it et al.} (1974) derived the KMS condition for thermal equilibrium states. In
the following we will establish related stability criteria for symbolic dynamics.

Consider a classical time-discrete, invertible dynamical system $(X, \Phi)$ given by a compact
Hausdorff space as its phase space $X$ and a map $\Phi: X \to X$. The flow of the system is generated by
the time iterates $\Phi^t$, $t \in \mathbb{Z}$, i.e., $t \mapsto \Phi^t$ is a one-parameter group for the
dynamics. Then, the function space of complex-valued continuous functions over $X$, $\mathfrak{A} = C(X)$,
yields a $\mathrm{C}^*$-algebra of classical observables for that dynamical system.

The states of such a $\mathrm{C}^*$-dynamical system are linear, positive, normalized functionals $\rho
: \mathfrak{A} \to \mathbb{C}$. For classical dynamical systems they correspond to probability measures
$\mu_\rho$ over the phase space $X$, such that $\rho(f) = \int_X f(x) \, d\mu_\rho(x)$ for $f \in C(X)$.
While pure states can be identified with single points in phase space $x \in X$, non-pure states are
statistical states given by measures $\mu_\rho$ that are not concentrated on a single point.

In its simplest sense, the stability of a dynamical system refers to the stability of a point $x^* \in X$
under the flow $\Phi^t$: $x^* = \Phi(x^*)$, i.e., $x^*$ is a fixed-point attractor. Limit-cycles or
higher-order tori as attractors can be related to fixed points by the technique of Poincar\'{e} sections. In
general, attractors are invariant sets $A \subset X$, such that $\Phi(A) \subseteq A$ and $\Phi^{-1}(A)
\subseteq A$. This invariance property of $A$ extends to probability measures $\mu$ according to
$\mu(\Phi^{-1}(A)) = \mu(A)$ which are called \emph{stationary} or \emph{invariant measures}. Likewise, a
statistical state $\rho_\mu$ over the algebra of continuous functions assigned to the measure $\mu$ has the
invariance property. The invariance of thermal equilibrium states is the first postulate by Haag {\it et al.}
(1974).

Structural stability refers to perturbations in the function space of the flow map $\Phi$. The system $(X,
\Phi)$ is called \emph{structurally stable} if there is a neighborhood $\mathcal{N}$ of $\Phi$ such that all
$\Psi \in \mathcal{N}$ are topologically equivalent with $\Phi$.\footnote{
Two maps $\Phi, \Psi$ are called \emph{topologically equivalent}, or \emph{conjugated}, if there is a
homeomorphism $h$ such that $h \circ \Phi = \Psi \circ h$. }
As Haag {\it et al.} (1974) pointed out, the concept of structural stability is closely related to that of
ergodicity. An invariant probability measure $\mu$ is said to be \emph{ergodic} under the flow $\Phi$ if an
invariant set $A$, has either measure zero or one: $\mu(A) \in \{0, 1\}$. If $\mu$ is non-ergodic, there is
an invariant set $A$ with $0 < \mu(A) < 1$ corresponding to an accidental degeneracy. Such
degeneracies are not stable under small perturbations. Hence, non-ergodic systems are in general not
structurally stable (Haag {\it et al.} 1974).

Thermal equilibrium states are given by invariant, ergodic measures over $X$. Beyond fixed points and limit
tori, more complicated attractors are mixing in addition. Mixing refers to the loss of temporal correlations
among the observables of a dynamical system. Formally, a measure $\mu$ is called \emph{mixing} if $|\mu(A
\cap \Phi^{-t}(B)) - \mu(A) \mu(B)| \stackrel{t \to \infty}{\longrightarrow} 0$ for all measurable sets $A,
B$ (Luzzatto preprint). This property can be rephrased by the correlation of observables $f, g \in
\mathfrak{A}$ at time $t$: $\mathcal{C}_t(f, g) = |\rho_\mu(f \cdot g \circ \Phi^t) - \rho_\mu(f) \cdot
\rho_\mu(g)|$ where $\rho_\mu$ is the statistical state assigned to the measure $\mu$. If
$\mathcal{C}_t(\chi_A, \chi_B) \stackrel{t \to \infty}{\longrightarrow} 0$ for characteristic functions
$\chi_A, \chi_B$ of the sets $A, B \subset X$, $\mu$ is mixing (Luzzatto, preprint). Interestingly, Haag
{\it et al.} (1974) derived this loss of correlations from a more fundamental, purely algebraic stability
property called ``asymptotic abelianness'' (Bratteli and Robinson 1997). The mixing property of a state
$\rho$ follows from the asymptotic abelianness of the algebra under the assumption that $\rho$ is
\emph{relatively pure}, i.e.~$\rho$ cannot be decomposed into a convex sum of invariant states ($\rho$ might
be decomposable into non-invariant states, however). Relatively pure states have sharp expectation values and
correspond, therefore, to thermodynamic macrostates (Shalizi and Moore preprint).

Stationarity (invariance), structural stability (ergodicity) and asymptotic abelianness (mixing) are
important for the investigation of nonlinear dynamical systems. Many rigorous results are known for
\emph{hyperbolic systems} where either the whole phase space possesses a hyperbolic structure (Anosov
diffeomorphisms) or there is a hyperbolic attractor. Anosov diffeomorphisms are known to be structurally
stable (Robinson 1999), and systems with hyperbolic attractors have invariant, ergodic and mixing probability
measures due to a theorem by Sinai, Ruelle and Bowen (Ruelle 1968, 1989). For non-hyperbolic systems, much
less is known (cf.~Viana {\it et al.} 2003).

Now let us introduce the notion of {\it epistemic} observables. For this purpose, consider a piecewise
constant function $f$ over the phase space $X$. Such a function is generally not overall continuous and does
therefore not belong to the $\mathrm{C}^*$-algebra $\mathfrak{A} = C(X)$ of observables. Instead, it belongs
to the larger $\mathrm{W}^*$-algebra\footnote{The relationship between $\mathrm{C}^*$- and
$\mathrm{W}^*$-algebras can be illustrated in the following way. Regarding a $\mathrm{C}^*$-algebra
$\mathfrak{A}$ as a complex vector space one can construct the dual $\mathfrak{A}^*$ of linear functionals
containing the states over $\mathfrak{A}$. This is again a vector space that becomes a Hilbert space in the
GNS construction and has a dual $\mathfrak{A}^{**}$. The original $\mathrm{C}^*$-algebra $\mathfrak{A}$ can
be canonically embedded in $\mathfrak{A}^{**}$ by $a(\rho) = \rho(a)$ where $\rho \in \mathfrak{A}^*$, the
right-hand side $a \in \mathfrak{A}$, and the left-hand side $a \in \mathfrak{A}^{**}$. Hence,
$\mathfrak{A}^{**}$ inherits the properties of $\mathfrak{A}$ (including the $\mathrm{C}^*$-property). The
fact that it has a Hilbert space as its predual turns it into a $\mathrm{W}^*$-algebra. The bidual
$\mathfrak{A}^{**}$ is generally much larger than $\mathfrak{A}$ and contains the epistemic observables.} of
$\hat{\mu}$-essentially bounded epistemic observables $L^\infty(X, \hat{\mu})$ that are contextually defined
by a reference probability measure $\hat{\mu}$ on the phase space $X$ used for a Gel'fand-Naimark-Segal (GNS)
construction (Primas 1998, Atmanspacher and Bishop, this issue). Two states $x, y \in X$ are called
\emph{epistemically equivalent with respect to $f$} if $f(x) = f(y)$ (beim Graben and Atmanspacher, in
press). Epistemically equivalent states are not distinguishable by means of the observable $f$. The classes
of epistemically equivalent states partition the phase space $X$ into disjoint sets $A_i$.

A finite partition of $X$, $\mathcal{P} = \{A_i | i \le I \}$, $A_i \cap A_j = \emptyset$ ($i \ne j$),
$\bigcup_i A_i = X$, also called a \emph{coarse-graining}, yields a symbolic dynamics (Lind and Marcus 1995)
of the system $(X, \Phi)$ in the following way: Taking the finite index set of the partition as an alphabet
$\mathbf{A}$ of cardinality $I$, one assigns to each initial condition $x_0 \in X$ a bi-infinite sequence $s
= \dots a_{i_{-1}} a_{i_0} . a_{i_1} a_{i_2}\dots$ of symbols $a_{i_k} \in \mathbf{A}$ according to the rule
$x_0 \mapsto s$, if $\Phi^t(x_0) \in A_{i_t}$, $t \in \mathbb{Z}$ (the dot indicates the origin of the time
scale). This mapping $s = \pi(x_0)$ is continuous in the topology of the space of sequences $\Sigma =
\mathbf{A}^\mathbb{Z}$. Accordingly, the first iterate $x_1 = \Phi(x_0)$ of $x_0$ is mapped onto the sequence
$s' = \dots a_{i_{-1}} a_{i_0} a_{i_1} . a_{i_2}\dots$. Therefore, the sequence $s'$ is obtained by shifting
all symbols of $s$ one place to the left.

A symbolic dynamical system is given by $(\Sigma, \sigma)$ where $\sigma(s) = s'$ is the left-shift. Since
the dynamics on $\Sigma$ is trivially represented by the shift $\sigma$, all important information is
now encoded in the structure of the symbolic sequences $s$. Therefore, symbolic dynamics deals with syntax
and pattern analysis (Lind and Marcus 1995, Keller and Wittfeld 2004, Steuer {\it et al.} 2004, Steuer {\it
et al.} 2001).

The systems $(X, \Phi)$ and $(\Sigma, \sigma)$ are related to each other by
\begin{equation}\label{intertwiner}
\pi \circ \Phi = \sigma \circ \pi \, ,
\end{equation}
which can be represented diagrammatically as:
\begin{center}
\unitlength 0.7mm \linethickness{0.4pt}
\begin{picture}(36.00,36.00)
\put(05.00,35.00){\line(1,0){22.00}}
\put(05.00,0.00){\line(1,0){22.00}}
\put(35.00,30.00){\line(0,-1){25.00}}
\put(00.00,30.00){\line(0,-1){25.00}}
\put(00.00,07.00){\vector(0,-1){2.0}}
\put(35.00,07.00){\vector(0,-1){2.0}}
\put(27.00,34.90){\vector(1,0){2.00}}
\put(27.00,-0.10){\vector(1,0){2.0}}
\put(00.00,0.00){\makebox(0,0)[cc]{$s$}}
\put(36.00,00.00){\makebox(0,0)[cc]{$\sigma(s)$}}
\put(00.00,35.00){\makebox(0,0)[cc]{$x$}}
\put(36.00,35.00){\makebox(0,0)[cc]{$\Phi(x)$}}
\put(04.00,18.00){\makebox(0,0)[cc]{$\pi$}}
\put(31.00,18.00){\makebox(0,0)[cc]{$\pi$}}
\put(17.50,03.00){\makebox(0,0)[cc]{$\sigma$}}
\put(17.50,31.00){\makebox(0,0)[cc]{$\Phi$}}
\end{picture}
\end{center}
where $\pi: X \to \Sigma$ acts as an \emph{intertwiner}. If $\pi$ is continuous and invertible and its
inverse $\pi^{-1}$ is also continuous, the maps $\Phi$ and $\sigma$ are topologically equivalent. In this
case the partition $\mathcal{P}$ is called {\it generating}. For {\it generating partitions}, the
correspondence between the phase space and the symbolic representation is one-to-one: each point in phase
space is uniquely represented by a bi-infinite symbolic sequence and \emph{vice versa}. Additionally, all
topological information is preserved.

Generating partitions are generally hard to find. However, it is known that hyperbolic systems possess
generating partitions for which the resulting symbolic dynamics is a Markov chain (Sinai 1968a,b, Bowen 1970).
The partitions that achieve this are so-called Markov partitions. The symbolic dynamics obtained from a
Markov partition is a \emph{shift of finite type}. This can be seen by defining an $I \times I$ stochastic
matrix $P_{ij} = \mu(\Phi^{-1}(A_i) \cap  A_j) / \mu(A_i)$ (Froyland 2001), where $\mu$ is a probability
measure. The associated transition matrix $T_{ij} = \sgn(P_{ij})$ provides a subset $\Sigma_T \subset \Sigma$
of admissible sequences. The sequence $s = \dots a_{i_{-1}} a_{i_0} . a_{i_1} a_{i_2}\dots$ belongs to
$\Sigma_T$ if $T_{a_{i_k} a_{i_{k+1}}} = 1$ (i.e. the transition from $a_{i_k}$ to $a_{i_{k+1}}$ is allowed).
The left-shift restricted to $\Sigma_T$ yields then a subshift of finite type $(\Sigma_T,
\sigma|_{\Sigma_T})$.

Assume that a coarse-grained description of a dynamical system $(X, \Phi)$ is such a shift of finite type
$(\Sigma_T, \sigma|_{\Sigma_T})$. Then we can distinguish two important cases. In the first case, either the
matrix $T$ itself or some power $T^l$ ($l > 1$) of $T$ is diagonal. If $T$ is diagonal, the cells $A_i$ of
the partition $\mathcal{P}$ are invariant sets under the flow $\Phi$. That is, the partition is coarse enough
to capture the asymptotically stable fixed points and limit tori together with their basins of attraction
 of a multistable dynamical system. Such systems are structurally stable unless they give rise
to bifurcations. If the $l$-th power of $T$ is diagonal, the admissible sequences of the symbolic dynamics
are periodic and $T$ is called {\it cyclic}. This means that the boundaries of the partition are
transversally intersected by a limit torus, which is asymptotically stable as well. The space of symbolic
sequences $\Sigma_T$ for these systems can be equipped with invariant, ergodic measures by taking Dirac
measures for the periodic sequences.

The second important case refers to an \emph{irreducible} transition matrix $T$, i.e.~there is a number $l$
such that $T^l$ is positive. Then the corresponding shift of finite type $(\Sigma_T, \sigma|_{\Sigma_T})$ is
an ergodic and mixing Markov chain where the eigenvector $p^*$ to eigenvalue one of the stochastic matrix $P$
corresponds to a unique invariant, ergodic measure that is mixing in addition to the first case above (Ruelle
1968, 1989). A well-elaborated theory relates these measures to KMS states in algebraic quantum statistics
(Olesen and Petersen 1978, Bratteli and Robinson 1997, Pinzari {\it et al.} 2000, Exel 2004), at least for
structurally stable hyperbolic systems. Such systems have Markov partitions enabling the construction of
thermal equilibrium KMS states (under certain conditions) which are also structurally stable (Robinson
1999). Furthermore, Markov partitions are generating and, thereby, admit a symbolic dynamics that is
topologically equivalent to the underlying phase space dynamics.

To conclude, subshifts of finite type $(\Sigma_T, \sigma|_{\Sigma_T})$, characterized by an $I \times
I$ transition matrix $T$, are structurally stable if $T$ is either cyclic (i.e.~there is an $l \ge 1$ such
that $T^l$ is diagonal) or if $T$ is irreducible. In both cases, the existence of invariant and ergodic
measures ensure stability conditions as required for the contextual emergence of epistemic observables and
associated states in a partitioned phase space.

\section{Contextual Emergence of Mental States \\ from Neural States}

Let us now consider a neurodynamical system $N = (X, \Phi)$ with phase space $X$ described by neural
observables $f_i: X \to \mathbb{R}$ (e.g.~spike rates or action potentials or somato-dendritic membrane
potentials of neurons) such that $x \in X$ is a point or, likewise, an \emph{activation vector} of a neural
population given by the values $(f_i(x))_{i \le n} \in \mathbb{R}^n$ for $n$ degrees of freedom. In the
following subsections we address three different ways of introducing epistemic observables on such a phase
space. The structural stability of their associated symbolic dynamics, which is of key significance for
contextual emergence, will be emphasized in particular.

\subsection{Neural Correlates of Consciousness}

There is a great variety of conscious mental states forming a \emph{mental state space} $Y$. Mental states
range from coarsest-grained (``just being conscious'') to finer-grained states such as wakefulness versus
sleep, dreaming, hypnosis, attentiveness, etc.\footnote{A recent empirically based study concerning the
relation between neural and mental state space representations for wakefulness versus sleep and other,
subtler examples (selective attention, intrinsic perceptual selection) is due to Fell (2004). For alternative
state space approaches see Wackermann (1999) and Hobson {\it et al.} (2000), and the following subsection.}
Even more refined are states of consciousness associated with specific {\it phenomenal content} (Chalmers
2000).

It is generally assumed that some neural system $N$ with phase space $X$ is correlated with particular mental
states $C \in Y$.  They can be related to epistemic observables $p:X \to \{0, 1\}$, where $p(x) = 1$ if the
activation vector $x$ is actually correlated with the mental state $C$. A \emph{phenomenal family}
$\mathcal{P} = \{ C_1, \dots C_I \}$ is a Boolean classification of pairwise disjoint states that cover the
whole mental state space $Y$ (Chalmers 2000). In other words, $\mathcal{P}$ provides a partition of the
mental state space $Y$ into $I$ states $C_i$. The whole mental state space can then be represented by a
system of such partitions of different coarse grainings: At the lowest level there is a binary partition
defining mental states of ``being conscious'' and ``not being conscious''. At subsequent levels, there are
more refined partitions defining, for instance, states of ``wakefulness'', ``sleep'', and altered states
(e.g.~hypnosis), again covering the entire mental state space $Y$.

According to Chalmers (2000), a neural correlate of consciousness (NCC) can be characterized by a minimal
sufficient neural subsystem $N$ that is correlated with a conscious state $C \in Y$. This characterization
refers to the interlevel relation of supervenience. The sufficiency of $N$ means that the activity of $N$
implies being in conscious state $C$.

From the point of view of this contribution, however, it is also appropriate to look for a necessary neural
subsystem $N$ whose activation is correlated with the conscious state $C$ in the sense of contextual
emergence. Being in a conscious state $C$ implies the activity of $N$, so that this activity is a necessary
condition for $C$.

Suppose that $N$ is an NCC for a conscious state $C_i \in \mathcal{P}$ with multiple realizations by
different activation patterns of $N$. Then different neural states, $x, y \in X$, are sufficient for the
conscious state $C_i$. Since $p_i(x) = p_i(y)$, $x$ and $y$ are epistemically indistinguishable from one
another and, hence, epistemically equivalent with respect to the observable $p_i$ corresponding to the mental
state $C_i \in \mathcal{P}$. In this sense, the partition $\mathcal{P}$ of the mental state space $Y$ induces
a partition $\mathcal{Q} = \{ A_1, \dots A_I \}$ of the neural state space $X$ into classes of epistemically
equivalent neural states. Labeling the cells $A_i$ of $\mathcal{Q}$ by symbols $a_i$ of a finite alphabet
$\mathbf{A}$, we obtain a symbolic representation of the mental states, emerging from the neural state space
by the mapping $\pi: \mathcal{P} \to \mathbf{A}$, $\pi(C_i) = a_i$. The dynamics of states $a_i$ in
$\mathbf{A}$ is a discrete sequence of symbols as a function of time, establishing a symbolic dynamics. If
the transitions between states of consciousness can be described by an $I \times I$ transition matrix $T$,
the mental symbolic dynamics is of finite type.

A coarse-grained partition of $X$ implies neighborhood relations between states in $Y$ that are different
from those in the underlying neural phase space $X$; in this sense it implies a change of topology. Also, 
the algebra of mental observables differs from that of neurobiological observables.
Obviously, these two differences depend essentially on the choice of the contextual partition of $Y$, based
on the choice of a phenomenal family, inducing the partition of $X$. We will now show that a particular 
concept of stability is crucial for a proper choice of such a partition and, thus, crucial for a properly conceived
relation between $X$ and $Y$.

The crucial demand for contextual emergence is that the equivalence classes of neural states in $X$ and,
hence, the mental states in $Y$ be structurally stable (in the sense of Sec.~2) under the dynamics in $X$.
Consider, e.g., the partition of $Y$ into the mental states ``wakefulness'' and ``sleep'' leading to two
disjoint sets in $X$. Given an appropriate discretization of time, the transition matrix $T$ is cyclic with
$T^2 = E$ ($E$ denoting the $2 \times 2$ unit matrix). That is, the coarse-grained description provides a
limit torus. By contrast, a sufficiently fine-grained partition of $Y$ into mental states of different
phenomenal content would have to be described by a high-dimensional irreducible transition matrix $T$ since
any such state should be connected to any other state by a symbolic trajectory of sufficient length. In this
case the resulting symbolic dynamics is an ergodic, mixing Markov chain with a distinguished KMS equilibrium
state (Pinzari {\it et al.} 2000, Exel 2004).

These stationary and structurally stable symbolic dynamical systems have strikingly different consequences
(beim Graben and Atmanspacher, in press). While fixed points and limit tori do not possess generating
partitions (beim Graben 2004), Markov chains can be obtained from Markov partitions which are generating.
Generating partitions admit a continuous approximation of individual points in the neural phase space $X$ by
symbolic sequences in $\mathbf{A}$ with arbitrary precision. Hence, the neural description in $X$ and the
coarse-grained, mental description in $Y$ are topologically equivalent.

This shows that the generating property of a partition is an important constraint for a viable symbolic
description of a system. Although this is a clear-cut criterion, generating partitions are notoriously
difficult to find in practice, and they are explicitly known for only very few examples. Nevertheless, they
are viable candidates for the implementation of a stability criterion appropriate for the contextual
emergence of mental states. A related stability constraint has been proposed recently (Werning and Maye
2004, this issue). An alternative approach, focusing on information constraints rather than stability, is
due to Shalizi and Moore (preprint).

\subsection{Macroscopic Neural States}

Another approach, leading to coarse-grained neural states without involving mental states is based on
mass potentials such as local field potentials (LFP) at the mesoscopic and the electroencephalogram (EEG) at
the macroscopic level of brain organization. Let $F: X \to \mathbb{R}$ be such an observable given by a mean
field
\begin{equation}\label{meanfield}
F(x) = \sum_i f_i(x) \ ,
\end{equation}
where the sum extends over a population of $n$ neurons and $f_i$ denotes a projector of $X$ onto the $i$-th
coordinate axis measuring the microscopic activation of the $i$-th neuron.

Similar to the previous subsection, the outcomes of $F$ have multiple realizations since the terms in the sum
in Eq.~(\ref{meanfield}) can be arranged arbitrarily. Therefore, two neural activation vectors $x, y$ can
lead to the same value $F(x) = F(y)$  (e.g. when $f_i(x) - \epsilon =  f_j(x) + \epsilon$, $i \ne j$), so
that they are indistinguishable by means of $F$ and, therefore, epistemically equivalent. If the equivalence
classes of $F$ in $X$ form a finite partition $\mathcal{Q} = \{ A_1, \dots A_I \}$ of $X$, we can again
assign symbols $a_i$ from an alphabet $\mathbf{A}$ to the cells $A_i$ and obtain a symbolic dynamics. In this
way, experimentally well-defined meso- and macroscopic brain observables, LFP and EEG, form a coarse-grained
description of the underlying microscopic neurodynamics. It should be emphasized that related approaches do
not involve any reference to concrete mental or conscious states. Whether or not one wants to relate
corresponding coarse-grained neural states to mental states is left open (for attempts in this
direction, see Fell (2004), Wackermann (1999) and Hobson {\it et al.} (2000)).

Coarse-grainings based on the symbolic encoding of EEG time series became increasingly popular in recent
years (Keller and Wittfeld 2004, beim Graben {\it et al.} 2000, Frisch {\it et al.} 2004, Frisch and beim
Graben 2005, Drenhaus {\it et al.}, in press, Schack 2004, Steuer {\it et al.} 2004). Since such partitions
are not induced by well-defined mental observables, it is unclear whether the stability conditions required
for contextual emergence are satisfied. It is, thus, particularly important to check this carefully.

One option to do this is to look for Markov partitions of the phase space which minimize correlations between
their cells, thus creating a Markov process for the symbolic dynamics of the meso- or macro- observables
if the dynamics in $X$ is
chaotic.\footnote{Evidence for chaotic brain processes has often been reported (cf.~Kaneko and Tsuda 2000,
and references therein).} Since Markov partitions are generating, they can be operationally identified by the
fact that the dynamical entropy for a generating partition is the supremum over all possible partitions, the
so-called Kolomogorov-Sinai entropy (see Atmanspacher (1997) for an annotated introduction). Iterative
partitioning algorithms in this and similar contexts have been discussed by Froyland (2001): Starting with an
initial partition, those sets which contribute to the greatest mass of the assumed invariant ergodic measure
are refined iteratively. Optimal partitions are thereby generated by a dynamics in ``partition space''.

Alternatively, the measured meso- or macroscopic observables can be analyzed by segmentation techniques
(Lehmann {\it et al.} 1987, Wackermann {\it et al.} 1993, Hobson {\it et al.} 2000, Hutt 2004, Schack
2004).\footnote{Lehmann {\it et al.} (1987) called the corresponding states ``brain microstates'' or ``atoms
of thought'', expressing the suggestion that they correspond to elementary ``chunks'' of consciousness.} A
recent proposal to implement this is due to Froyland (2005): One tries to partition the space $X$ into
\emph{almost invariant sets} such that trajectories spend most of the time within individual cells of the
partition, and transitions between cells are likely at larger time scales. In this way, the dynamics on short
time scales is described by cyclic transition matrices, whereas large time scales yield descriptions by
Markov processes with irreducible transition matrices. The separation of time scales provides, then, a
contextual criterion for properly defined macroscopic brain states.

\subsection{Remarks on Symbol Grounding}

The \emph{symbol grounding problem} posed by Harnad (1990) refers to the problem of assigning meaning to
symbols on purely syntactic grounds, as proposed by cognitivists such as Fodor and Pylyshin (1988). This
entails the question of how conscious mental states with phenomenal content can be characterized by their
NCC. Chalmers (2000) defined an NCC for phenomenal content as a neural system $N$ with ``systematicity in the
correlation'', meaning that the representation of a content in $N$ is correlated with a representation of
that content in consciousness. In other words, there should be a mapping from the neural state space $X$ onto
the space of conscious states $Y$ such that regions in $X$ are related to phenomenal contents in $Y$. This
mapping differs from the mapping required for contextual emergence as discussed in Sec.~3.1. For a neural
representation of content further constraints are crucial.

First, all states in $B \subset X$ representing the same content $C$ should be similar in some respect: there
should be a mapping $g: X \to X$, let us call it a \emph{gauge transformation}, such that $B$ is invariant
under $g$, $g(B) \subset B$. In this sense, $g$ is a {\it similarity transformation}. On the other hand,
graded differences in phenomenal similarity should be reflected by topologically neighboring regions in phase
space. One would, therefore, require the mapping $g$ to be a homeomorphism, leading to \emph{topographic
mappings} of contents (Chalmers 2000).

A second requirement is the \emph{compositionality} of representations (Fodor and Pylyshin 1988, Werning and
Maye 2004). Compositionality refers to the relation between syntax and semantics insofar as the meaning of a
composed (or ``complex'') symbol is a function of the meanings of its constituent symbols and the way
they are put together. A prerequisite for compositionality is the existence of syntactic rules determining
which composites are \emph{constituents} of a language and which are not. (In our approach, constituents are
admissible (sub-)sequences in the corresponding symbolic dynamics (beim Graben 2004).)

According to Harnad, these constraints need to be combined with his proposal that symbols
must be grounded in embodied cognition. They represent objects or facts from the environments
of physically embodied agents that collect information by their sensory apparatuses and act by their motor
effectors. While Harnad (1990) suggests a hybrid architecture consisting of a neural network as an
invariance detector and a classical symbol processor to meet the compositionality constraint, we shall
discuss the alternative of a unified neurodynamical system.

This can be achieved using the notion of \emph{conceptual spaces} as discussed by G\"ardenfors (2004). A
conceptual space is a vector space spanned by quantitative observables. The conceptual space for color, e.g.,
can be constructed as the three-dimensional RGB coordinate system or an equivalent representation
supplied by the cones in the retina (Steels and Belpaeme in press). According to G\"ardenfors (2004) a
\emph{natural concept} is then a convex region in a conceptual space such that all elements in that region
are similar in a particular context. Implementing conceptual spaces by neural systems, we arrive again at
partitions of neurodynamical phase spaces. The idea of  gauge invariance yields partitions of finest grain,
corresponding to ``natural kinds'' (Carnap 1928/2003, Quine 1969), that might be too refined for other
contexts.

Such contexts can be supplied by pragmatic accounts. Suppose a toy-world in which only orange objects are
eatable, and all other objects are not (Steels and Belpaeme in press). Then, a binary partition of color
space into ``orange'' and ``non-orange'' will be sufficient for an agent to survive. Thus, survival (or
successful communication) serve as contextual constraints for the emergence of cognitive symbols. Symbol
grounding corresponds then to \emph{categorization} of conceptual spaces driven by pragmatic goals.

The contextual emergence of symbols in partitioned conceptual spaces raises the question of the stability of
the symbols. The dynamics that has to be taken into account now is, however, not neurodynamics but rather
sociodynamics: the evolution of populations of cognitive agents. (Neurodynamically, concepts are static
objects given by the cells of a partition.) An interesting approach in this sense has been developed within
the framework of \emph{evolutionary game theory} (Steels and Belpaeme, in press, J\"ager 2004, van Rooy 2004).
In these models the phase space is spanned by the population numbers of agents with competing strategies. The
outcome of the games is assessed by a utility function which in turn determines the number of offspring
of the players. In cognitive applications of evolutionary game theory, offspring means adoption of the
winning strategy by other players.

If categories or concepts are given by partitions of conceptual spaces, competing strategies are different
partitions of the same local conceptual spaces shared by different agents. Evolutionary game
theory then describes a dynamics in partition space similar to the search for optimal partitions by iterative
algorithms (Froyland 2001). Evolutionary stable strategies are asymptotically stable fixed points in
evolutionary game theory (J\"ager 2004). This stability criterion means that cultural evolution grounds
symbols in shared partitions of local conceptual spaces of cognitive agents.

The structural stability of dynamically evolving partitions can be illustrated by the ``naming game'' (van
Rooy 2004). When a categorization in conceptual space is fixed the cells are labeled by symbols of an
alphabet $\mathbf{A}$. This can be done arbitrarily by convention, or it can be achieved by another pragmatic
game that optimizes the utility reward. For instance, assume that two meanings $m_1, m_2$ assign two symbolic
forms $f_1, f_2$, that $m_1$ is less complex than $m_2$, and the same for the forms $f_1, f_2$. For such a
scenario, van Rooy (2004) found only two evolutionarily stable strategies: the \emph{Horn strategy} which
assigns more complex forms to more complex meanings, and the \emph{anti-Horn strategy} performing otherwise.
Since the basin of attraction of the Horn strategy is larger than that of the anti-Horn strategy (J\"ager
2004), the Horn strategy provides a higher degree of structural stability.

\section{Compatibility of Psychological Descriptions}

It is an old and much discussed question whether and, if yes, how psychology could become a unified science,
integrating the many approaches and models that constitute its contemporary situation. It is often argued
that the largely fragmented appearance of psychology (and cognitive science as well) is due to the fact that
psychology is still in a preparadigmatic, ``immature'' state. Some have even argued that this situation is
unavoidable (e.g., Koch 1983, Gardner 1992) and should be considered as the strength of psychology (e.g.,
Viney 1989, McNally 1992) rather than an undesirable affair.

From the perspective of the philosophy of mind, arguments against the possibility of a unified science of
psychology have been presented as well. Most prominent are the accounts of Kim (1992, 1993) and Fodor (1997),
both using the scheme of multiple realization in the framework of supervenience to reject unification.
Shapiro (in press) has recently pointed out particular weak points in their arguments.

On the other hand, there is a growing interest in articulating visions for a unified science of psychology,
cognition, or consciousness (see, e.g., Newell 1990, Anderson 1996). Recently, various approaches have been
proposed to reach a degree of coherence comparable to established sciences as, e.g., physics with
well-defined relations between its different disciplines. Examples are approaches such as ``psychological
behaviorism'' (Staats 1996, 1999), ``unified psychology'' (Sternberg and Grigorenko 2001, Sternberg {\it et
al.} 2001), and the ``tree of knowledge system'' (Henriques 2003). A key feature in the latter program is the
{\it commensurability} of competing approaches in psychology, explicated by Yanchar and Slife (1997) and
Slife (2000).

This section presents a way in which the notion of commensurable models can be implemented in a formally
rigorous fashion. A suitable way to formulate commensurability in technical terms is related to the concept
of {\it compatibility}. Briefly speaking, two models are considered as commensurable if they are compatible
in the sense that there exist well-defined mappings between them. If this is not the case, they are
incompatible. It turns out that the scheme of contextual emergence provides some detailed and clarifying
insights how to proceed in this regard. The two levels of description whose interlevel relations are
significant for this purpose are those of neurobiology and psychology or cognitive science, respectively.
Compatible and incompatible implementations of cognitive symbol systems have recently been discussed by beim
Graben (2004).

A key result of the work by beim Graben and Atmanspacher (in press) is that a non-generating partition is
incompatible with any other partition (even if this is generating) in the sense that there is no well-defined
mapping between the partitions.\footnote{Two partitions ${\cal P}_1$ and ${\cal P}_2$ are (in)compatible if
their $\sigma$-algebras are (not) identical up to $\mu$-measure zero. Two partitions are maximally
incompatible, or complementary, if their $\sigma$-algebras are disjoint up to the entire phase space $X$
(cf.~beim Graben and Atmanspacher, in press).} As a consequence, models based on such partitions are
incompatible as well. Since any {\it ad hoc} chosen partition is quite unlikely to be generating, it may be
suspected that the resulting incompatibility of models based on such partitions is the rule rather than the
exception. While incompatibility may admit the possibility of ``partially coherent'' models, the case of
maximal incompatibility, also called complementarity, excludes any coherence between different models
completely.

At this point it should be clear that our notion of incompatibility is more subtle than a ``logical
incompatibility'' (Slife 2000) in the sense that two models are simply negations of each other. Also, it
would be interesting to compare Slife's (2000) complementary models with our formal approach in terms of
maximally incompatible models, which are basically incoherent only in a Boolean framework. From a perspective
admitting non-Boolean descriptions, the notion of coherence acquires a more comprehensive meaning, including
complementary descriptions as representations of an underlying, more general description (see Primas 1977).

With these remarks in mind, incompatible models due to non-generating partitions represent a significant
limit to the vision of a unified or integrative science of psychology. Or, turned positively, such a
unification will be strongly facilitated if the approaches to be unified are based on generating, hence
compatible, partitions that are structurally stable and induced by well-defined mental or cognitive states.
As mentioned, it is a tedious task to identify such generating partitions. Nevertheless, the necessary formal
and numerical tools are available today and significantly facilitated by the symbolic description using
shifts of finite type and transition matrices. All one has to do is find transitions between mental states
that are irreducible, yielding a stationary, ergodic, and mixing Markov chain with distinguished KMS states.

If there is a good deal of empirical plausibility for a particular partition, one might hope that this
implies that such a partition is generating (at least in an approximate sense) and, thus, that the
corresponding mental or cognitive states are stable (in the sense of the KMS condition). However,
there may be cases of conflict between the empirical and the theoretical constraint on a proper partition. In
such cases, one has to face the possibility that the ``empirical plausibility'' of cognitive states may be
unjustified, e.g.~based on questionable prejudices. If cognitive states turn out to be dynamically unstable,
this theoretical argument against their adequacy is very strong indeed.

Compatible partitions and, consequently, compatible psychological models show another important feature that
is occasionally addressed in current literature: the topological equivalence of representations in
neurodynamic and mental state spaces (cf.~Metzinger 2003, p.~619, and Fell (2004) for empirically based
examples). Topological equivalence ensures that the mapping between $X$ and $Y$ is faithful in the sense that
the two state space representations yield equivalent information about the system (see Sec.~2).
Non-generating, incompatible partitions do not provide representations in $Y$ that are topologically
equivalent with the underlying representation in $X$.

\section{Summary}

The relation between mental states and neural states is discussed in the framework of a recently proposed
scheme of interlevel relations called contextual emergence. According to this proposal, knowledge of the
neural description provides necessary but not sufficient conditions for a proper psychological description.
Sufficient conditions can be defined as contingent contexts at the cognitive (phenomenal) level and
implemented as stability criteria at the underlying neural level.

This procedure has been demonstrated using the terminology of symbolic dynamics at the cognitive level.
Equivalence classes of neural states are defined as neural correlates of mental states represented
symbolically. Mental states are well-defined if criteria of temporal and structural stability are satisfied
for their neural correlates. These criteria can be implemented either by generating or, more specifically,
Markov partitions; or by partitions of asymptotically stable fixed points or limit tori. This implies that
proper mental or cognitive states must satisfy appropriate stability conditions.

If this is not explicitly taken care of for chaotic systems admitting generating partitions, one has to
expect that \emph{ad hoc} selected partitions are not generating. As a consequence, models based on such
partitions are incompatible. This may be a possible source of the long-standing problem of how to develop a
unified science of psychology. Only for carefully chosen generating partitions it can be guaranteed that
different cognitive models are compatible and, hence, can have transparent relations with respect to each
other.

Moreover, psychological (or cognitive) models are topologically equivalent with their neurobiological basis
only if they are constructed from generating partitions. Without cognitive contexts serving as sufficient
conditions for compatibility and topological equivalence, the neurobiological level of description provides
only necessary conditions for psychological descriptions.

\section*{Acknowledgments}

We are grateful to Jiri Wackermann and three anonymous referees for their helpful suggestions how to
improve an earlier version of this paper.

\section*{References}

\begin{description}

\item Anderson, J. A., and Rosenfeld, E. (1989), \textit{Neurocomputing: Foundations of Research}. Cambridge:
MIT Press.

\item Anderson, N. (1996), {\it A Functional Theory of Cognition}, Mahwah: Erlbaum.

\item Atmanspacher, H. (1997), ``Dynamical Entropy in Dynamical Systems'', in H. Atmanspacher and E. Ruhnau
(eds.), {\it Time, Temporality, Now}. Berlin: Springer, pp. 327--346.

\item Atmanspacher, H., and Bishop, R. (this issue), ``Stability Conditions in Contextual Emergence,''
{\em Chaos and Complexity Letters}.

\item Bishop, R. C., and Atmanspacher, H. (preprint), ``Contextual Emergence in the Description of Properties''.

\item Bowen, R. (1970), ``Markov Partitions for Axiom A Diffeomorphisms,'' {\it Am.~J.~Math.} 92: 725--747.

\item Bratteli, O. and Robinson, D.L. (1997), {\it Operator Algebras and Quantum Statistical Mechanics 2},
Berlin: Springer.

\item Carnap, R. (1928/2003), {\it The Logical Structure of the World and Pseudoproblems in Philosophy},
Peru (IL): Open Court.

\item Chalmers, D. (2000), ``What Is a Neural Correlate of Consciousness?'', in T.~Metzinger (ed.), {\it
Neural Correlates of Consciousness}, Cambridge: MIT Press, pp.~17--39.

\item Drenhaus, H., beim Graben, P., Saddy, D., and Frisch, S. (in press), \enquote{Diagnosis and Repair of
Negative Polarity Constructions in the Light  of Symbolic Resonance Analysis}, {\em Brain and Language}.

\item Exel, R. (2004), \enquote{{KMS} states for generalized gauge actions on {C}untz-{K}rieger algebras
({A}n application of the {R}uelle-{P}erron-{F}robenius theorem)}, {\em Bull. Brazil. Math. Soc.} 35(1):
1 -- 12.

\item Fell, J. (2004), ``Identifying Neural Correlates of Consciousness: The State Space Approach'', {\it
Consciousness and Cognition} 13: 709--729.

\item Fodor, J. (1997), ``Special Sciences: Still Autonomous After All These Years'', {\it Philosophical
Perspectives} 11: 149--163.

\item Fodor, J. and Pylyshyn, Z.~W. (1988) \enquote{Connectionism and Cognitive Architecture: A Critical
Analysis}, {\em Cognition} 28: 3 -- 71.

\item Frisch, S. and beim Graben, P. (2005), \enquote{Finding Needles in Haystacks: Symbolic Resonance
Analysis of Event-Related Potentials Unveils Different Processing Demands}, {\em Cogn. Brain Res.}, 24(3):  476 -- 491.

\item Frisch, S., beim Graben, P., and Schlesewsky, M. (2004), \enquote{Parallelizing Grammatical
Functions: {P600} and {P345} Reflect Different Cost of Reanalysis} {\em Int. J. Bifurcation Chaos}, 14(2): 531
-- 551.

\item Froyland, G. (2001), ``Extracting Dynamical Behavior Via Markov Models,'' in A.I.~Mees (ed.), {\it
Nonlinear Dynamics and Statistics}, Boston: Birkh\"auser pp.~281--312.

\item Froyland, G. (2005), \enquote{Statistically Optimal Almost-Invariant Sets}, {\em Physica D} 200:
205 -- 219.

\item G\"ardenfors, P. (2004) \enquote{Conceptual Spaces as a Framework for Knowledge
  Representations}, {\em Mind and Matter} 2(2): 9 -- 27.

\item Gardner, H. (1992), ``Scientific Psychology: Should We Bury It or Praise It?'' {\it New Ideas in
Psychology} 10: 179--190.

\item beim Graben, P. (2004), ``Incompatible Implementations of Physical Symbol Systems'', {\it Mind and
Matter} 2(2): 29--51.

\item beim Graben, P., and Atmanspacher, H. (in press), ``Complementarity in Classical Dynamical Systems'',
{\it Found.~Phys.}

\item beim Graben, P., Saddy, J.D., Schlesewsky, M., and Kurths, J. (2000), ``Symbolic Dynamics of
Event-Related Brain Potentials,'' {\em Phys.~Rev.~E}, 62: 5518--5541.

\item Haag, R. Kastler, , and Trych-Pohlmeyer, E.~B. (1974), ``Stability and Equilibrium States,'' {\em
Commun. Math. Phys.} 38: 173--193.

\item Harnad, S. (1990), ``The Symbol Grounding Problem,'' {\it Physica D} 42: 335--346.

\item Henriques, G.R. (2003), ``The Tree of Knowledge System and the Theoretical Unification of Psychology'',
{\it Review of General Psychology} 7: 150--182.

\item Hobson, J.A., Pace-Schott, E.F., and Stickgold, R. (2000), ``Dreaming and the Brain: Toward a Cognitive
Neuroscience of Conscious States,'' {\it Behavioral and Brain Sciences} 23: 793--842.

\item Hutt, A. (2004), \enquote{An Analytical Framework for Modeling Evoked and Event-Related
Potentials}, {\em Int. J. Bifurcation Chaos}, 14(2): 653 -- 666.

\item J\"ager, G. (2004), {\it Evolutionary Game Theory for Linguists: A Primer}. Unpublished manuscript,
Stanford University and University of Potsdam.

\item Kaneko, K., and Tsuda, I. (2000), \textit{Complex Systems: Chaos and Beyond}. Berlin: Springer.

\item Keller, K., and Wittfeld, K. (2004), ``Distances of Time Series Components by Means of Symbolic
Dynamics,'' {\it Int.~J.~Bif.~Chaos} 14: 693--704.

\item Kim, J. (1992), ``Multiple Realization and the Metaphysics of Reduction'', {\it Philosophy and
Phenomenological Research} 52: 1--26.

\item Kim, J. (1993), \textit{Supervenience and Mind}. Cambridge: Cambridge University Press.

\item Koch, S. (1993), `` `Psychology' or `the Psychological Studies'?'' {\it American Psychologist} 48:
902--904.

\item Lehmann, D., Ozaki, H., and Pal, I. (1987), ``EEG Alpha Map Series; Brain Micro-States by
Space-Oriented Adaptive Segmentation,'' {\it Electroencephalogr.~Clin.~Neurophysiol.} 67: 271--288.

\item Lind, D., and Marcus, B. (1995), {\it Symbolic Dynamics and Coding}, Cambridge: Cambridge University
Press.

\item Luzzatto, S. (preprint), ``Stochastic-like Behaviour in Nonuniformly Expanding Maps,''
arXiv:mathDS/0409085.

\item McNally, R.J. (1992), ``Disunity in Psychology: Chaos or Speciation?'' {\it American Psychologist} 47:
1054.

\item Metzinger, T. (2003), {\it Being No One}, Cambridge: MIT Press. \item Newell, A. (1990), {\it Unified
Theories of Cognition}, Cambridge: Harvard University Press.

\item Olesen, D. and Petersen, G.K. (1978), ``Some C$^*$-Dynamical Systems With a Single KMS State,'' {\it
Math.~Scand.} 42: 111--118.

\item Pinzari, C., Watatani, Y. and Yonetani, K. (2000), ``KMS States, Entropy and the Variational Principle
in Full C$^*$-Dynamical Systems,'' {\it Commun.~Math.~Phys.} 213: 331--379.

\item Primas, H. (1977), ``Theory Reduction and Non-Boolean Theories,'' {\it Journal of Mathematical Biology}
4: 281--301.

\item Primas, H. (1998), ``Emergence in Exact Natural Sciences,''\ \textit{Acta Polytechnica Scandinavica}
91: 83--98.

\item Quine, W.~V. (1969) \enquote{Natural Kinds}, in {\em Ontological Relativity and
  Other Essays}, New York: Columbia University Press.

\item Robinson, C. (1999), {\em Dynamical Systems. Stability, Symbolic Dynamics, and Chaos}, Boca Raton:
CRC Press.

\item van~Rooy, R. (2004) \enquote{Signalling Games Select {H}orn Strategies}, {\em  Linguistics and
Philosophy} 27: 491 -- 527.

\item Ruelle, D. (1968), ``Statistical Mechanics of a One-Dimensional Lattice Gas,'' {\it
Commun.~Math.~Phys.} 9: 267--278.

\item Ruelle, D. (1989), ``The Thermodynamic Formalism for Expanding Maps,''
{\it Commun.~Math.~Phys.} 125: 239--262.

\item Schack, B. (2004), \enquote{How to Construct a Microstate-Based Alphabet for Evaluating Information
Processing in Time}, {\em Int. J. Bifurcation Chaos} 14(2): 793 -- 814.

\item Shalizi, C.R., and Moore, C. (preprint), ``What Is a Macrostate? Subjective Observations and Objective Dynamics''.

\item Shapiro, L. (in press), ``Can Psychology Be a Unified Science?'' {\it Philosophy of Science}.

\item Sinai, Ya.G. (1968a), ``Markov Partitions and C-Diffeomorphisms,''
{\it Functional Analysis and Its Applications} 2: 61--82.

\item Sinai, Ya.G. (1968b), ``Construction of Markov Partitions,''
{\it Functional Analysis and Its Applications} 2: 245--253.

\item Slife, B. (2000), ``Are Discourse Communities Incommensurable in a Fragmented Psychology?''
{\it Journal of Mind and Behavior} 21: 261--271.

\item Staats, A.W. (1996), {\it Behavior and Psychology: Psychological Behaviorism}, New York: Plenum.

\item Staats, A.W. (1999), ``Uniting Psychology Requires New Infrastructure, Theory, Method, and a
Research Agenda'', {\it Review of General Psychology} 3: 3--13.

\item Steels, L. and Belpaeme, T. (in press), \enquote{Coordinating Perceptually Grounded Categories Through
Language: A Case Study for Colour}, {\it Behavioral and Brain Sciences}.

\item Sternberg, R.J., and Grigorenko, E.L. (2001), ``Unified Psychology'',
{\it American Psychologist} 56: 1069--1079.

\item Sternberg, R.J., Grigorenko, E.L., and Kalmar, D. (2001), ``The Role of Theory in Unified Psychology'',
{\it Theoretical and Philosophical Psychology} 21: 99--117.

\item Steuer, R., Ebeling, W., Bengner, T., Dehnicke, C., H\"attig, H., and Meencke, H.-J. (2004),
\enquote{Entropy and Complexity Analysis of Intracranially Recorded EEG}, {\em Int. J. Bifurcation Chaos}
14(2): 815--824.

\item Steuer, R., Ebeling, W., Russel, D., Bahar,S., Neiman, A., and Moss., F. (2001), ``Entropy and
Local Uncertainty of Data from Sensory Neurons,'' {\em Phys. Rev. E} 64: 061911.

\item Viana, R.L., Grebogi, C., de S.~Pinto, S.E., and Barbosa, J.R.R. (2003), ``Pseudo-deterministic Chaotic
Systems,'' {\em Int. J. Bifurcation Chaos} 13: 1--19.

\item Viney, W. (1989), ``The Cyclops and the Twelve-Eyed Toad: William James and the
Unity--Disunity Problem in Psychology'', {\it American Psychologist} 44: 1261--1265.

\item Wackermann, J., Lehmann, D., Michel, C.M., and Strik, W.K. (1993), ``Adaptive Segmentation
of Spontaneous EEG Map Series into Spatially Defined Microstates,''  {\it International Journal
of Psychophysiology} 14: 269--283

\item Wackermann, J. (1999), ``Towards a Quantitative Characterisation of Functional
States of the Brain: From the Non-Linear Methodology to the Global Linear Description.
{\it International Journal of Psychophysiology} 34: 65--80.

\item Werning, M. and Maye, A. (2004), ``Implementing the (De-)Compositionality of Concepts: Oscillatory
Networks, Coherency Chains and Hierarchical Binding,'' in S.D.~Levy and R.~Gayer (eds.), {\em Compositional
Connectionism in Cognitive Science}, Menlo Park: AAAI Press, pp.~67--81.

\item Werning, M. and Maye, A. (this issue), ``The Cortical Implementation of Complex Attribute and
Substance Concepts: Synchrony, Frames, and Hierarchical Binding,'' {\em Chaos and Complexity Letters}.

\item Yanchar, S.C., and Slife, B.D. (1997), ``Pursuing Unity in a Fragmented Psychology: Problems and
Prospects'', {\it Review of General Psychology} 1: 235--255.

\end{description}

\end{document}